\documentclass[preprint,showpacs,titlepage,aps,prd,tightenlines,amsmath,byrevtex,nofootinbib]{revtex4}
\usepackage{graphicx}

\begin{document}

\title{
Search for sterile neutrinos at reactors
}

\author{Osamu~Yasuda}
\email{E-mail: yasuda_at_phys.se.tmu.ac.jp}
\affiliation{Department of Physics, Tokyo Metropolitan University,
Hachioji, Tokyo 192-0397, Japan}


\vglue 1.4cm
\begin{abstract}
The sensitivity to the sterile neutrino mixing
at very short baseline reactor neutrino experiments
is investigated.  In the case of 
conventional (thermal neutron) reactors
it is found that the sensitivity is lost for
$\Delta m^2 \gtrsim$ 1 eV$^2$ due to
smearing of the reactor core size.
On the other hand, in the case of
an experimental fast neutron reactor Joyo,
because of its small size,
sensitivity to $\sin^22\theta_{14}$
can be as good as 0.03 for
$\Delta m^2 \sim$ several eV$^2$
with the Bugey-like detector setup.
\end{abstract}


\maketitle

\section{Introduction}

Since the LSND group announced the anomaly which suggests neutrino
oscillations with mass squared difference of ${\cal
O}$(1) eV$^2$\cite{Athanassopoulos:1996jb,Athanassopoulos:1997pv,Aguilar:2001ty},
schemes with sterile neutrinos have attracted a lot of attention.
This is because the standard three flavor scheme has only two
independent mass squared differences, i.e., $\Delta m^2_{21}=\Delta
m^2_\odot\simeq 8\times 10^{-5}$eV$^2$ for the solar neutrino
oscillation, and $|\Delta m^2_{31}|=\Delta m^2_{\text{atm}}\simeq
2.4\times 10^{-3}$eV$^2$ for the atmospheric neutrino oscillation, and
it does not have room for the mass squared difference of ${\cal
O}$(1) eV$^2$.  The extra state has to be sterile neutrino, which is
singlet with respect to the gauge group of the Standard Model,
because the number of weakly interacting light neutrinos has to be
three from the LEP data.\cite{Nakamura:2010zzi}

The LSND anomaly has been tested by the MiniBooNE experiment.
While the MiniBooNE data on the neutrino mode~\cite{AguilarArevalo:2007it}
disfavors the region suggested by LSND, their data
on the anti-neutrino mode~\cite{AguilarArevalo:2010wv}
seems to be consistent with the LSND data.

Recently the flux of the reactor neutrino was recalculated
in Ref.~\cite{Mueller:2011nm} and it was reported that the
normalization is shifted by about +3\% on
average.\footnote{ The increase in the flux claimed by
Ref.~\cite{Mueller:2011nm} was confirmed by an independent
calculation in Ref.~\cite{Huber:2011wv}.}  Then a
re-analysis of 19 reactor neutrino results at short
baselines~\cite{Mention:2011rk} in the light of new
reactor neutrino spectra~\cite{Mueller:2011nm} may
indicate neutrino oscillation at $\Delta
m^2\gtrsim$1eV$^2$.

Reactor experiments with more than one detector have attracted much
attention as a possibility to measure $\theta_{13}$
precisely~\cite{Kozlov:2001jv,Minakata:2002jv,Huber:2003pm,Anderson:2004pk},
and three experiments~\cite{Ardellier:2004ui,Ahn:2010vy,Guo:2007ug}
are now either running or expected to start soon.
In the three flavor case with
$|\Delta m^2_{31}|=2.4\times10^{-3}$eV$^2$,
it was shown assuming infinite statistics
that the optimized baseline lengths $L_F$ and $L_N$ for the far
and near detectors are $L_F\simeq$1.8km and $L_N\simeq$0km in the rate
analysis~\cite{Yasuda:2004dd,Sugiyama:2004bv}, while they are
$L_F\simeq$10.6km and $L_N\simeq$8.4km in the spectrum
analysis~\cite{Sugiyama:2005ir}.
Unfortunately, in order to justify the assumption
on negligible statistical errors for $L\sim$10km,
one would need unrealistically huge detectors,
so one is forced to choose the baseline
lengths which are optimized for the rate analysis
for $\Delta m^2=2.4\times10^{-3}$eV$^2$.
On the other hand, if we perform an oscillation
experiment to probe $\Delta m^2\sim{\cal O}$(1) eV$^2$,
it becomes realistic to place the detectors at the baseline lengths
which are optimized for the spectrum analysis~(See Sect.
4 in the published version of Ref.~\cite{Sugiyama:2005ir}).

In this paper we discuss the sensitivity of very short
line reactor experiments to the sterile neutrino mixing
for $\Delta m^2\sim{\cal O}$(1) eV$^2$ in the so-called
(3+1)-scheme.\footnote{According to
Ref.~\cite{Kopp:2011qd}, the so-called
(3+2)-scheme~\cite{Sorel:2003hf} gives a better fit to the
global data than the (3+1)-scheme does.  However, we are
mainly interested here in the potential sensitivity to the
sterile neutrino mixing for $\Delta m^2\sim{\cal O}$(1)
eV$^2$, and for simplicity we will discuss the
(3+1)-scheme, leaving the analysis of the (3+2)-scheme for
the future.}  Proposals have been made to test the
bound of the Bugey reactor
experiment~\cite{Declais:1994su} on the sterile neutrino
mixing angle using a reactor~\cite{nucifer},\footnote{
See, e.g., Refs.~\cite{Sugiyama:2005ir} (the published
version), \cite{Latimer:2007qe,deGouvea:2008qk} for
earlier works on search for sterile neutrinos at a
reactor.} an
accelerator~\cite{Agarwalla:2009em,Rubbia:2011}, and a
$\beta$-source~\cite{Cribier:2011fv}.

Throughout this paper we discuss the case with
a single reactor and two detectors, which have
an advantage to cancel the systematic errors
that are correlated between detectors.
The conditions of the detectors are assumed to be
the same as those of the Bugey experiment.

In Sect.~2 we briefly review the four neutrino schemes.
In Sect.~3 we evaluate the sensitivity of very short line reactor
experiments to the sterile neutrino mixing.
In Sect.~4 we summarize our results.
In Appendix~\ref{appendix1} we give some details
to get an analytical expression for $\chi^2$.

\section{Four neutrino schemes}
\label{sec:schemes}

Four-neutrino schemes consist of one extra sterile state in addition
to the three weakly interacting ones.
Depending on whether one or two
mass eigenstate(s) are separated from the others by the largest
mass-squared gap, the schemes are
called (3+1)- and (2+2)-schemes.
The (2+2) schemes are excluded by the solar and
atmospheric neutrino data~\cite{Maltoni:2004ei}, so
we will not discuss the (2+2) schemes in this paper.
In the (3+1) schemes, on the other hand,
the phenomenology of solar and atmospheric oscillations is
approximately the same as that of the three flavor framework,
so as far as the tension
between the solar and atmospheric constraints
are concerned, the (3+1) schemes do not have any problem.
However, if we try to account for LSND and all other negative results
of the short baseline experiments, then the (3+1) schemes
have a problem.
To explain the LSND data while satisfying the
constraints from other disappearance experiments, the oscillation
probabilities of the appearance and disappearance channels have to
satisfy the following relation \cite{Okada:1996kw,Bilenky:1996rw}:
\begin{eqnarray}
\sin^22\theta_{\mbox{\rm\tiny LSND}}(\Delta m^2)
<\frac{1}{4}\,\sin^22\theta_{\mbox{\rm\scriptsize Bugey}}(\Delta m^2)
\cdot
\sin^22\theta_{\mbox{\rm\tiny CDHSW}}(\Delta m^2)
\label{relation31}
\end{eqnarray}
where $\theta_{\mbox{\rm\tiny LSND}}(\Delta m^2)$,
$\theta_{\mbox{\rm\tiny CDHSW}}(\Delta m^2)$,
$\theta_{\mbox{\rm\scriptsize Bugey}}(\Delta m^2)$ are the value of
the effective two-flavor mixing angle as a function of the mass
squared difference $\Delta m^2$ in the allowed region for LSND
($\bar{\nu}_\mu\rightarrow\bar{\nu}_e$), the CDHSW experiment
\cite{Dydak:1983zq} ($\nu_\mu\rightarrow\nu_\mu$), and the Bugey
experiment \cite{Declais:1994su}
($\bar{\nu}_e\rightarrow\bar{\nu}_e$), respectively.  The reason that
the (3+1)-scheme to explain LSND has been disfavored until
Refs.~\cite{Mueller:2011nm,Mention:2011rk} appeared is because
Eq.~(\ref{relation31}) is not satisfied for any value of $\Delta m^2$,
if we adopt the allowed regions in Refs.~\cite{Dydak:1983zq} and
\cite{Declais:1994su}.
However, if the flux of the reactor neutrino is slightly
larger than the one used in the Bugey
analysis~\cite{Declais:1994su}, the allowed region
becomes slightly wider and we have more chance to satisfy
Eq.~(\ref{relation31}).

In this paper we will use the following
parametrization for the mixing matrix,
adopted in Ref.~\cite{Maltoni:2007zf}:
\begin{eqnarray}
    U =
    R_{34}(\theta_{34} ,\, 0) \; R_{24}(\theta_{24} ,\, 0) \;
    R_{23}(\theta_{23} ,\, \delta_3) \;
    R_{14}(\theta_{14} ,\, 0) \; R_{13}(\theta_{13} ,\, \delta_2) \; 
    R_{12}(\theta_{12} ,\, \delta_1) \,,
    \nonumber
\end{eqnarray}
where $R_{jk}(\theta_{jk},\ \delta_l)$ are the complex
rotation matrices in the $jk$-plane defined as:
\begin{eqnarray}
[R_{jk}(\theta_{jk},\ \delta_{l})]_{pq} = 
\delta_{pq}+(\cos \theta_{jk}-1)
(\delta_{jp}\delta_{jq}+\delta_{kp}\delta_{kq})
+\sin \theta_{jk}
(e^{-i\delta_l}\delta_{jp}\delta_{kq}
-e^{i\delta_l}\delta_{jq}\delta_{kp}).
\nonumber
\end{eqnarray}
With this parametrization, for the
very short baseline reactor experiments,
where the average neutrino energy $E$ is
approximately 4MeV and the baseline length is
about 10m, we have $|\Delta m^2_{jk}L/4E|\ll 1~(j,k=1,2,3)$,
so that the disappearance probability
is given by 
\begin{eqnarray}
    P(\bar{\nu}_e\to\bar{\nu}_e) =
    1-\sin^22\theta_{14}\,\sin^2
\left(\frac{\Delta m^2_{41}L}{4E}
\right)
    \label{prob}
\end{eqnarray}
to a good approximation.
Eq.~(\ref{prob}) is the formula which will be
used in the oscillation analysis throughout this paper.

\section{Sensitivity to $\sin^2{2\theta_{14}}$
by a spectral analysis}

Throughout this paper we discuss the case with
a single reactor and two detectors.
The detectors are assumed to be of the Bugey type,
i.e., liquid scintillation detector of volume 600 liters
with the detection efficiency which yields
about 90,000 events at $L$=15m from a reactor of
a power 2.8GW after running for 1800 hours.
Also for simplicity, we assume in this paper that
the near and far detectors are identical and have the
same sizes of systematic errors.

\begin{figure}
\hspace*{-20mm}
\includegraphics[scale=0.8]{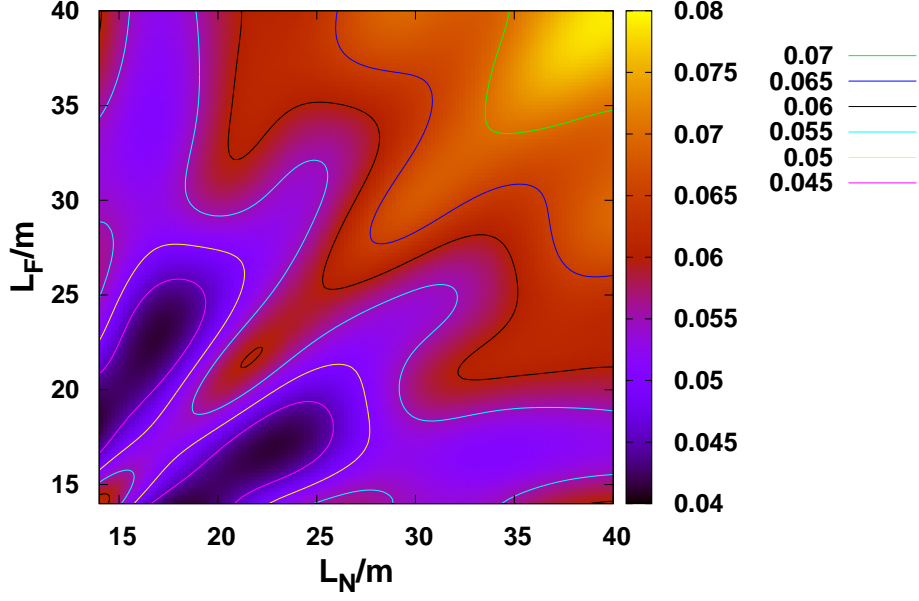}
\hspace*{-20mm}
\includegraphics[scale=0.8]{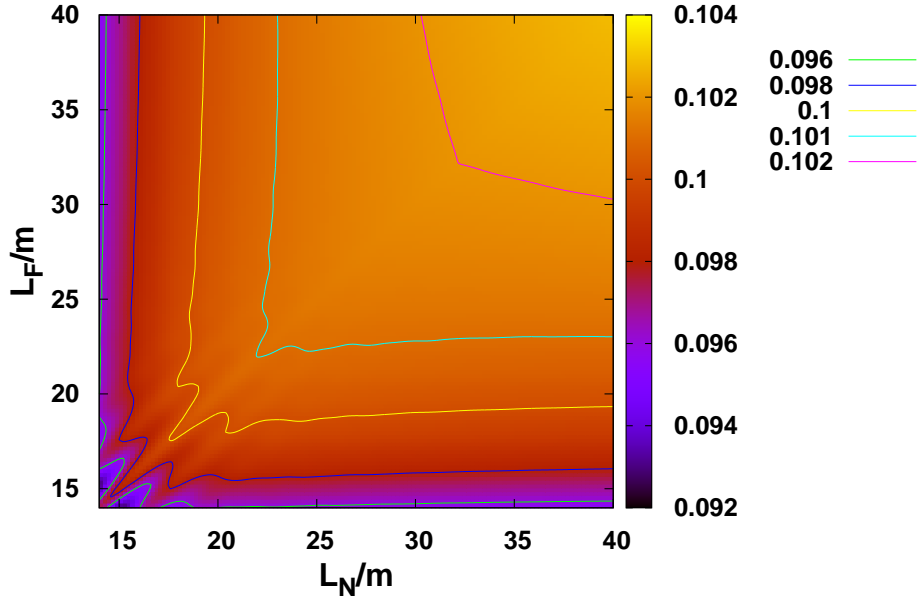}
\caption{
The sensitivity to $\sin^22\theta_{14}$
as a function of the two baseline lengths
in the case of an ordinary thermal neutron
reactor of a diameter 4m,
a height 4m, a thermal power 2.8GW.
$\Delta m^2_{41}=1$eV$^2$ (upper panel), 5eV$^2$ (lower) is assumed.
}
\label{fig1}
\end{figure}

To evaluate the sensitivity to $\sin^2{2\theta_{14}}$,
we introduce the following
$\chi^2$ which was adopted in Ref.~\cite{Sugiyama:2005ir}:
\begin{eqnarray}
\hspace*{-20mm}
\displaystyle
\chi^2&=&\min_{\alpha's}\Bigg\{
\displaystyle\sum_{A=N,F}\sum_{i=1}^n
\frac{1} {(t^A_i\sigma^A_i)^2}
\left[m^A_i-t^A_i(1+\alpha+\alpha^A+\alpha_i)
-\alpha_{\text{cal}}^A t^A_iv^A_i
\right]^2\nonumber\\
&+&\displaystyle\sum_{A=N,F}\left[\left(
\frac{\alpha^A} {\sigma_{\text{dB}}}\right)^2
+\left(\frac{\alpha_{\text{cal}}^A} {\sigma_{\text{cal}}}\right)^2\right]
+\displaystyle\sum_{i=1}^n\left(
\frac{\alpha_i} {\sigma_{\text{Db}}}\right)^2
+\left(\frac{\alpha} {\sigma_{\text{DB}}}\right)^2\Bigg\}.
\label{chipull}
\end{eqnarray}
Here, $m^A_i$ is the number of events to be measured
at the near ($A=N$) and far ($A=F$)
for the $i$-th energy bin
with the neutrino oscillation\footnote{
As we will see below,
in the spectrum analysis,
unlike in the rate one, it is not necessary for
the near detector to have events without oscillation,
since the dominant term in $\chi^2$ looks at the
difference between the maximum and minimum of the
oscillation pattern.  In fact, in Fig.~\ref{fig1} below
there are regions in which $L_N \ge L_F$.
Therefore, it may be misleading
to call the detector at the shorter baseline length
the near detector.  Nevertheless, we use these
words throughout this paper for simplicity.},
and $t^A_i$ is the theoretical prediction without the oscillation.
$(\sigma^A_i)^2$ is the uncorrelated error
which consists of the statistical plus uncorrelated bin-to-bin
systematic error:
\begin{eqnarray}
\hspace*{-20mm}
(t^A_i\sigma^A_i)^2
=t^A_i+\left(t^A_i\sigma^A_{\text{db}}\right)^2,
\nonumber
\end{eqnarray}
where $\sigma^A_{\text{db}}$ is the uncorrelated bin-to-bin
systematic error.
For simplicity
we assume that sizes of the bin-to-bin uncorrelated systematic errors
of the detectors and of the flux are independent of the energy.
Also we take the choice of the bins in such a way that
the number of events for each bin is equal: $t^A_i=t^A~(i=1,\cdots,n)$.
$\alpha$ is a variable which corresponds to a
common overall normalization error $\sigma_{\text{DB}}$ for
the number of events.  $\alpha^A~(A=N,F)$ is a variable
which introduces the detector-specific
uncertainties $\sigma_{\text{dB}}$ of the near and far detectors.
$\alpha_i~(i=1,\cdots,n)$ is a variable for
an uncertainty $\sigma_{\text{Db}}$ of the
theoretical prediction for each energy bin which
is uncorrelated between different energy bins.\footnote{
Here we follow the notation for the systematic errors in
Ref.~\cite{Sugiyama:2005ir}.
The first suffix of $\sigma$ stands for the property for the systematic error with respect
to the detectors while the second is with respect
to bins, and capital (small) letter stands for a correlated (uncorrelated) systematic error.
The correspondence for the notation in Ref.~\cite{Huber:2003pm} is as follows:
$\sigma_{\text{db}}=\sigma_{\text{exp}}$,
$\sigma_{\text{dB}}=\sigma_b$,
$\sigma_{\text{Db}}=\sigma_{\text{shape}}$,
$\sigma_{\text{DB}}=\sigma_a$.}
$\alpha_{\text{cal}}^A~(A=N,F)$ is a variable which introduces 
an energy calibration uncertainty $\sigma_{\text{cal}}$
and comes in the theoretical prediction in the form
of $(1+\alpha_{\text{cal}}^A)E$ instead of the observed energy $E$.
Thus, the deviation $v^A_i$ (divided by
the expected number of events) from the
theoretical prediction $t^A_i$ due to
this uncertainty can be written as
\begin{eqnarray}
\hspace*{-2mm}
v^A_i=\lim_{\alpha_{\text{cal}}^A\rightarrow0}
\frac{1}{\alpha_{\text{cal}}^A t^A_i}
\left[\frac{N_p T}{4\pi L_A^2}\int
dE\int_{(1+\alpha_{\text{cal}}^A)E_i}^{(1+\alpha_{\text{cal}}^A)E_{i+1}}dE'
\,R(E_e,E') \epsilon(E) F(E)\sigma(E)
-t^A_i\right],
\label{v}
\end{eqnarray}
with
\begin{eqnarray}
\hspace*{-20mm}
t^A_i\equiv\frac{N_p T}{4\pi L_A^2}\int
dE\int_{E_i}^{E_{i+1}}dE'
\,R(E_e,E') \epsilon(E) F(E)\sigma(E).
\label{t}
\end{eqnarray}
In Eqs.~(\ref{v}) and (\ref{t}),
$N_p$ is the number of target protons in the detector,
$T$ denotes the exposure time,
$L_A$ is the baseline for the detector $A$,
$F(E)$ is the flux of $\bar{\nu}_e$,
and $\sigma(E)$ is the cross section of the
inverse $\beta$ decay $\bar{\nu}_e+p\to n+e^+$.
$E$ is the energy of the incident $\bar{\nu}_e$
and it is related to the positron energy $E_e$
and the masses $m_n$, $m_p$ of a neutron and a proton by
$E=E_e+m_n-m_p=E_e+1.3$ MeV.  $E'$ is the
measured positron energy and we will assume
that the energy resolution is given by $8\%/\sqrt{E}$,
i.e., $R(E_e,E')=R(E-m_n+m_p,E')$ is a Gaussian
function which describes the energy resolution and is given by
$R(E_e,E')=(1/\sqrt{2\pi}\sigma)\exp[-(E_e-E')^2/2\sigma^2]$,
where $\sigma
=0.08\sqrt{(E_e+m_e)/\mbox{MeV}}
=0.08\sqrt{(E-0.8/\mbox{MeV})/\mbox{MeV}}$.
Our strategy is to
{\it assume no oscillation in the theoretical prediction} $t^A_i$,
to substitute the number of events {\it with oscillation} in $m^A_i$,
and to see the sensitivity by looking at the value of
$\chi^2$.  Since $\chi^2$ is quadratical in the variables
$\alpha$, $\alpha^A$, $\alpha_i$, $\alpha_{\text{cal}}^A$, we can minimize
with respect to these variables in Eq.~(\ref{chipull}) exactly.\footnote{
In principle we could take a different convention,
such as multiplying the measured numbers $m^A_i$ by
the uncertainty $(1+\alpha+\alpha^A+\alpha_i)$, etc., and in fact it is
done in some references.
With such a convention, it becomes complicated
to work with an analytical approach because $\chi^2$ is not a Gaussian
with respect to $m^A_i$, which includes oscillation parameters,
after the minimizations of $\alpha$'s.
However, the difference between such a
convention and ours affects only the higher orders
in $\sigma^2$'s.
The conclusion on the sensitivity
with such a convention should coincide with ours numerically.}
Since we assume
no oscillation for the theoretical prediction $t^A_i$,
the quantity
$v^A_i$ in Eq.~(\ref{v}) is independent of $A(=N,F)$:
\begin{eqnarray}
\hspace*{-20mm}
v^N_i=v^F_i=v_i.
\label{v3}
\end{eqnarray}

Here we will take the following assumptions for the systematic errors:
\begin{eqnarray}
\sigma_{\text{db}}&=&0.5\%,\nonumber\\
\sigma_{\text{dB}}&=&0.5\%,\nonumber\\
\sigma_{\text{Db}}&=&2\%,\nonumber\\
\sigma_{\text{DB}}&=&3\%,\nonumber\\
\sigma_{\text{cal}}&=&0.6\%.
\label{error}
\end{eqnarray}
We will take the number $n$ of bins $n$=32 and the energy
interval 2.8MeV$\le E_\nu\le$7.8MeV.
The measurement is supposed to continue for 1,800 hours,
and the total numbers of events is expected to be
approximately 90,000$\times (15\mbox{m}/L_A)^2~(A=N,F)$.

\subsection{An analytical expression for $\chi^2$}
As was demonstrated in
Refs.~\cite{Sugiyama:2004bv,Sugiyama:2005ir}, it is
instructive to have an analytic expression for $\chi^2$,
since it enables us to see which term becomes dominant in
the spectrum analysis.  Unfortunately, it is difficult to
evaluate $\chi^2$ analytically with non-zero statistical
errors and the energy calibration uncertainty
$\sigma_{\text{cal}}$.  In the limit of infinite
statistic, an analytical expression for $\chi^2$ can be
obtained (cf.  Eq.~(10) in Ref.~\cite{Sugiyama:2005ir}).
On the other hand, in the presence of finite statistical
errors (with equal numbers of events for each bin,
$t^A_j=t^A$), we can obtain an analytical form in the
limit of $\sigma_{\text{cal}}=0$ and the result is given
by Eq.~(\ref{chi6}) in Appendix~\ref{appendix1}.

In the present case, under
the assumption that the total numbers of events
$N^F$ and $N^N$ are of order $10^5$, that
the number $n$ of bins is 32, and that
the systematic errors are given by
Eq.~(\ref{error}), we get
\begin{eqnarray}
\hspace*{-35mm}
&{\ }&\frac{\chi^2}{\sin^42\theta_{14}}
\nonumber\\
&\simeq&
\frac{\left\{\vec{u}_1\cdot(\vec{D}^{F}
+\vec{D}^{N})\right\}^2/n}
{4\sigma^2_{\text{DB}}}
+\,\frac{
\left\{\vec{u}_1\cdot(\vec{D}^{N}
-\vec{D}^{F})\right\}^2/n}
{2\sigma^2_{\text{dB}}
+1/N^F+1/N^N
}
\nonumber\\
&{\ }&
+\sum_{j=2}^n\frac{
\left\{\vec{u}_j\cdot(c_2\vec{D}^{F}
+s_2\vec{D}^{N})\right\}^2/n}
{\sigma^2_{\text{Db}}/n
+(1/N^F+1/N^N)/2
+\left\{(1/N^F-1/N^N)^2/4
+(\sigma^2_{\text{Db}}/n)^2
\right\}^{1/2}
}
\nonumber\\
&{\ }&
+\sum_{j=2}^n\frac{
\left\{\vec{u}_j\cdot(c_2\vec{D}^{N}
-s_2\vec{D}^{F})\right\}^2/n}
{\sigma^2_{\text{Db}}/n
+(1/N^F+1/N^N)/2
-\left\{(1/N^F-1/N^N)^2/4
+(\sigma^2_{\text{Db}}/n)^2
\right\}^{1/2}
}.
\label{chi6app}
\end{eqnarray}
where we have introduced the variables
\begin{eqnarray}
\hspace*{-20mm}
D^A_i&\equiv&
-\frac{1}{\sin^22\theta_{14}}\frac{m^A_i-t^A_i} {t^A_i}
\nonumber\\
&=&
\frac{\displaystyle
\int dE\int_{E_i}^{E_{i+1}}
dE'~R(E_e,E')\epsilon(E)F(E)\sigma(E)\sin^2\left(
\frac{\Delta m^2_{41}L_A}{4E}
\right)}
{\displaystyle
\int dE\int_{E_i}^{E_{i+1}} dE'~R(E_e,E')\epsilon(E)F(E)\sigma(E)},
\end{eqnarray}
and $\vec{u}_j$ are the orthonormal eigenvectors
of some $n\times n$ unitary matrices (see Appendix~\ref{appendix1}).
$n$ is the number of bins and
$N^A~(A=N,F)$ stands for the total number of events
($n\times t_j^A=N^A$).
$c_2\equiv\cos\varphi_2$, $s_2\equiv\sin\varphi_2$,
and $\varphi_2$ are angles defined in
Eq.~(\ref{phi}) to diagonalize a $2\times2$ matrix.
Because $\sigma_{\text{Db}}^2/n \sim 1/(2N^F)$,
we have $0<\varphi_2\lesssim\pi/4$.
The numerators in all the four terms on RHS in
Eq.~(\ref{chi6app}) are divided by the number $n$
of bins, because it is known in the case of infinite
statistics~\cite{Sugiyama:2005ir} that
$\{\vec{u}_1\cdot(\vec{D}^F\pm\vec{D}^N)\}^2/n$ etc.
are almost independent of $n$ for $n\gtrsim 16$.

The first (last) two terms in Eq.~(\ref{chi6app}) correspond to
the rate (spectral) analysis, because
$\vec{u}_1\propto (1,\cdots,1)^T$, while the
other vectors $\vec{u}_j~(j=2,\cdots,n)$ are orthogonal to $\vec{u}_1$.
The first (second) term looks at the sum (difference) of the total numbers
of events at the far and near detectors.
The dominant error in the first term
is the systematic error $\sigma_{\text{DB}}^2$
which is correlated between the two detectors,
while the dominant one in the second
is the statistical error $1/N^F+1/N^N$
plus the systematic one $\sigma_{\text{dB}}^2$
which is uncorrelated between the two detectors.
These first two terms are basically what
was discussed in the rate analysis to
study the sensitivity to $\sin^22\theta_{13}$
in Ref.~\cite{Minakata:2002jv}.
The third (fourth) term examines the sum
(difference) of the significance of the spectrum shapes
of events at the far and near detectors.
As statistic increase,
the denominator of the fourth term becomes the smallest,
and it is the fourth term which dominates $\chi^2$
in the limit of infinite statistics.
With the finite numbers of events, however,
the last three terms do contribute to $\chi^2$.

In the limit of
large mass squared difference $\Delta m^2_{41}$, 
the vector $D^A_j$ becomes $D^A_j\to(1/2)(1,\cdots,1)$,
$\vec{u}_1\cdot\vec{D}^{A}\to\sqrt{n}/2$ and
$\vec{u}_j\cdot\vec{D}^{A}\to0~(j=2,\cdots,n)$.
So the right hand side of Eq.~(\ref{chi6app})
becomes $(4\sigma^2_{\text{DB}})^{-1}$, and the
sensitivity in this limit is given by
$\sin^22\theta_{14}\sim\sqrt{4\sigma^2_{\text{DB}}\chi^2}$
= $\sqrt{4\times 2.7}\times$~0.03 $\sim 0.1$. 
Note that $\chi^2 = 2.7$ gives a 90\%CL bound
on $\sin^22\theta_{14}$.

\begin{figure}
\vspace*{20mm}
\hspace*{-15mm}
\includegraphics[scale=0.8]{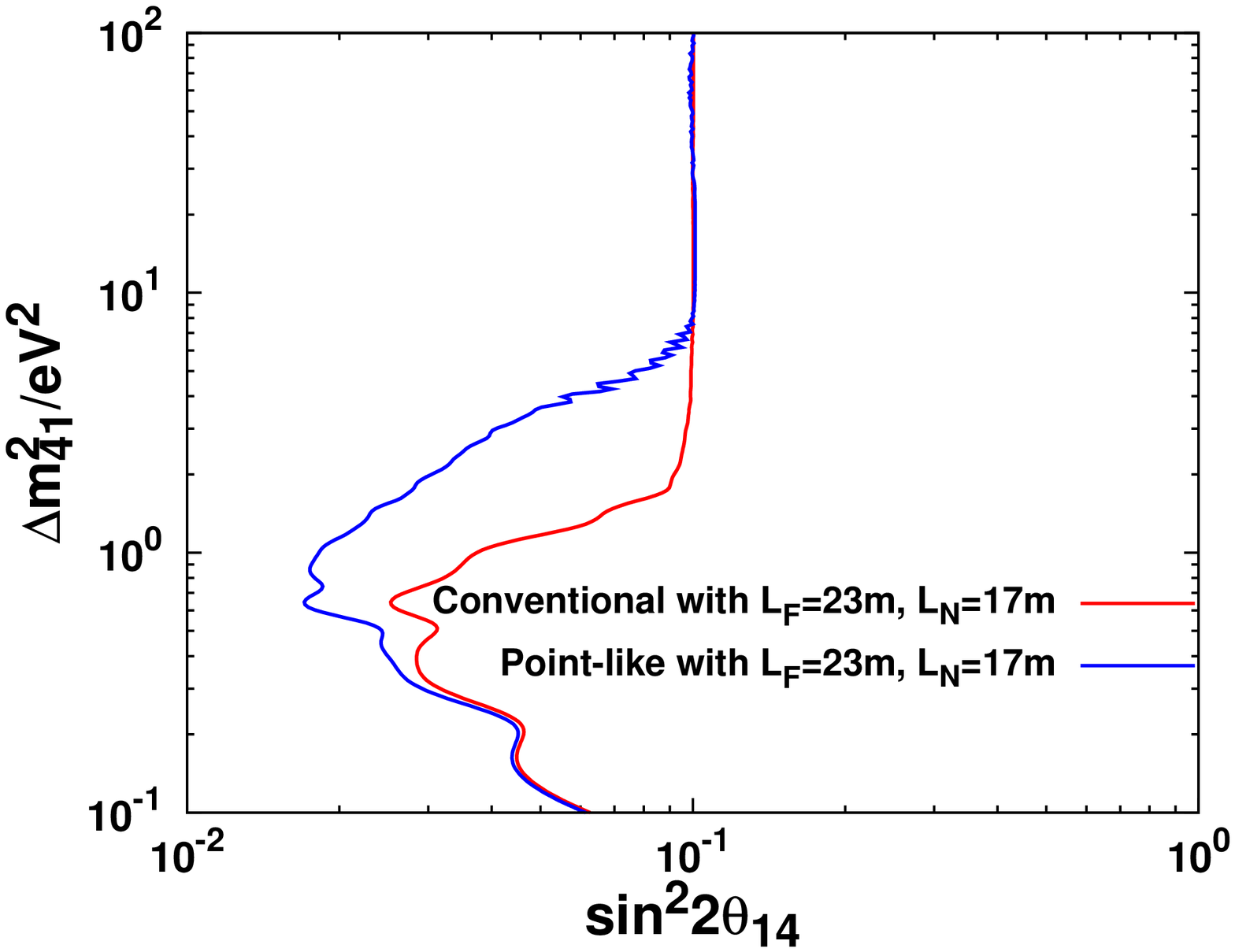}
\vglue 0.5cm
\caption{
The sensitivity to $\sin^22\theta_{14}$.
The red line stands for the one for
a thermal neutron reactor (of a diameter 4m,
a height 4m, a thermal power 2.8GW) with two
detectors at $L_F=23$m and $L_N=17$m,
while the blue line for the one for
a hypothetical reactor with a point-like
core.
}
\label{fig2}
\end{figure}

\subsection{Thermal neutron reactors}
A core of conventional (thermal neutron) reactors
typically has a diameter of 4m, a height 4m
and its thermal power is typically a few GW.
As we will see below, when we try to perform
an oscillation experiment at very short
baseline lengths, smearing of the reactor
core size gives a nontrivial contribution
to the results because the baseline lengths
are comparable to the core size.

We have computed $\chi^2$ in Eq.~(\ref{chipull})
numerically in the case of a thermal neutron reactor.
Fig.~\ref{fig1} shows the sensitivity
to $\sin^2{2\theta_{14}}$ as a function of
the baseline lengths of the near
and far detectors for $\Delta m^2_{41}=1$eV$^2$,
and 5eV$^2$, respectively.
We have assumed a thermal power of 2.8GW,
and a diameter of 4m and a height of 4m, and
we have varied the baseline lengths
in the range 14m$\lesssim L_A \lesssim $40m.
Unlike the case of infinite statistics~\cite{Sugiyama:2005ir},
the statistical errors are important in the
present setup of the detectors, and
longer baseline lengths are disfavored.

For $\Delta m^2_{41}=1$eV$^2$
(Fig.~\ref{fig1} upper panel),
the set $(L_N, L_F)\simeq$ (17m, 23m)
gives the optimum.  In contrast to the
rate analysis, in which the optimized
baseline length of the near detector
is $L_N$=0m to avoid oscillations,
the spectrum analysis with
$(L_N, L_F)=$ (17m, 23m) looks
at the difference between the
maximum and minimum of the spectrum shape
with neutrino oscillations
at $L_N$ and $L_F$
mainly for the energy region $E_\nu\sim$ 4MeV
where the number of events are expected to be
the largest.

On the other hand, for
$\Delta m^2_{41}=5$eV$^2$
(Fig.~\ref{fig1} lower panel),
the sensitivity to $\sin^2{2\theta_{14}}$
is as low as 0.1 for most of the set
of the baseline lengths $(L_N, L_F)$.
This implies that $\Delta m^2_{41}$ is
so large that we have average over rapid
oscillations for almost any baseline lengths
and the only term which contributes in
Eq.~(\ref{chi6app}) is the first term,
that corresponds to the rate analysis.

Fig.~\ref{fig2} shows the sensitivity
to $\sin^2{2\theta_{14}}$ as a function of
$\Delta m^2_{41}$ in the case of the baseline
lengths $(L_N, L_F)=$ (17m, 23m).
For $\Delta m^2_{41}\gtrsim 2$eV$^2$,
it indicates that the sensitivity is
no better than 0.1, which is basically
the result of the rate analysis.
The sensitivity in the case of a hypothetical
point-like reactor, where all the conditions
for the detectors are the same, is also given in
Fig.~\ref{fig2} for comparison.
Fig.~\ref{fig2} indicates that the sensitivity
would be as good as several $\times 10^{-2}$
for a few eV$^2$, if the core were point-like.
So we can conclude that we have poor sensitivity
for $\Delta m^2_{41}\gtrsim$ 2eV$^2$ because of
the smearing effect of the
finite core size of the reactor.
We have computed the sensitivity also for
different numbers of bins ($n$=16, 64), and
verified explicitly that the result of Fig.~\ref{fig2} 
is almost independent of $n$ for $16\le n\le 64$.
This is because the statistical errors are
dominant in $\chi^2$.

\begin{figure}
\hspace*{-20mm}
\includegraphics[scale=0.8]{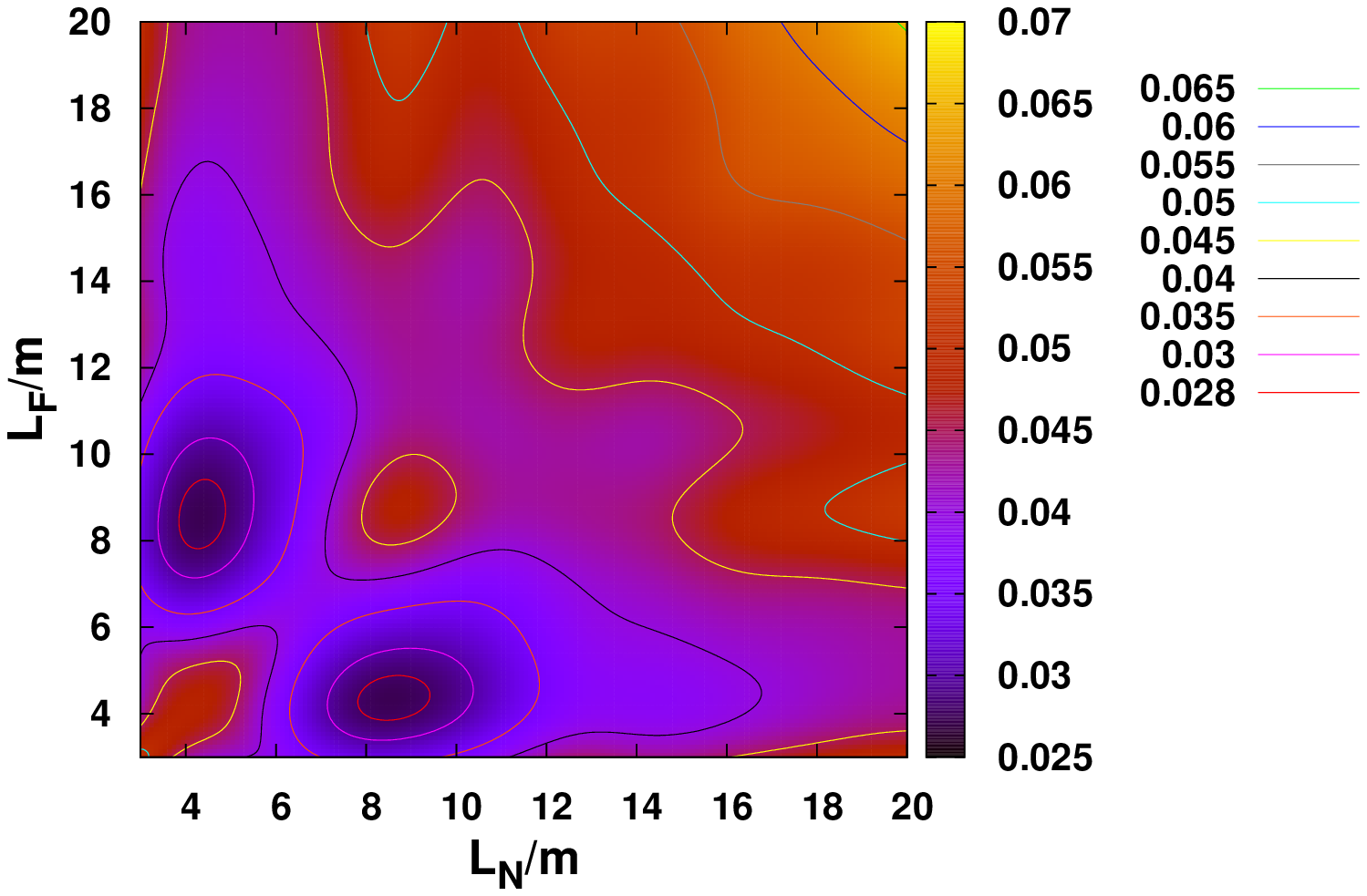}
\hspace*{-20mm}
\includegraphics[scale=0.8]{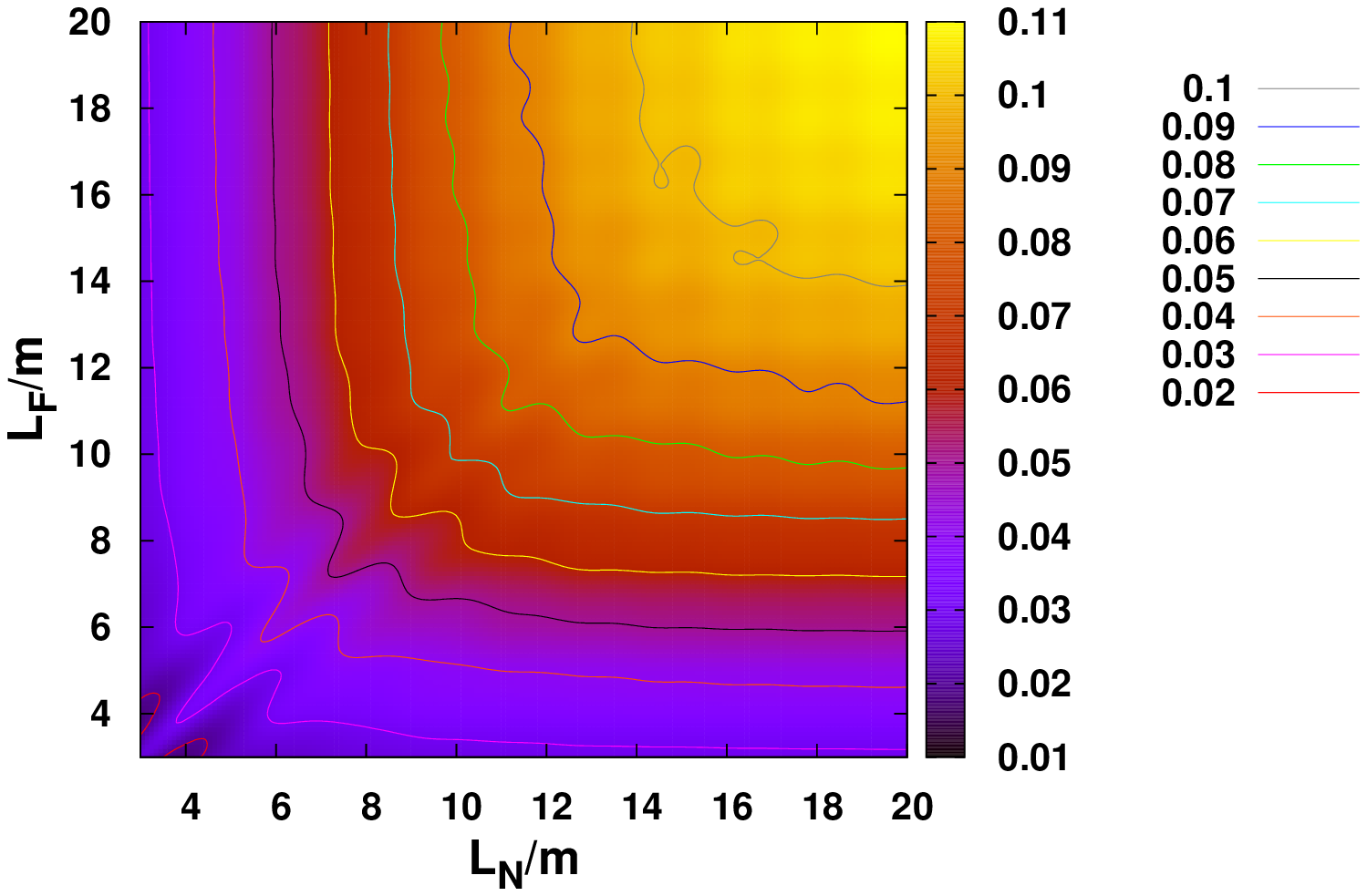}
\vglue 0.5cm
\caption{
The sensitivity to $\sin^22\theta_{14}$
as a function of the two baseline lengths
in the case of Joyo~\cite{joyo} with MK-III upgrade~\cite{joyomk3}
which is an experimental fast reactor
(of a diameter 0.8m,
a height 0.5m, a thermal power 0.14GW).
$\Delta m^2_{41}=1$eV$^2$ (upper panel), 5eV$^2$ (lower) is assumed.
}
\label{fig3}
\end{figure}

\begin{figure}
\vspace*{20mm}
\hspace*{-15mm}
\includegraphics[scale=0.8]{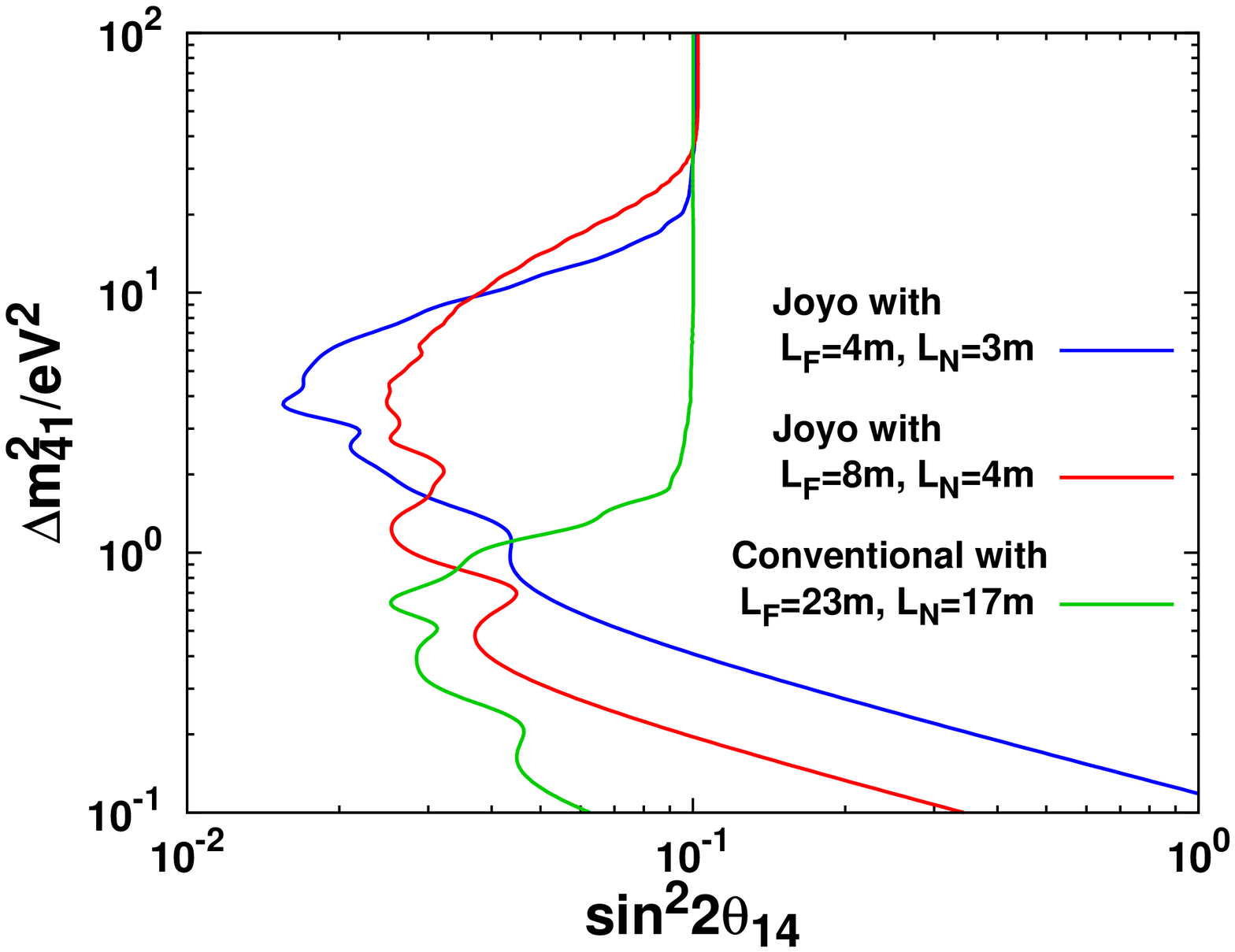}
\vglue 0.5cm
\caption{
The sensitivity to $\sin^22\theta_{14}$
of the Joyo reactor with two detectors.
The Blue line stands for the one in the
case with two detectors at $L_F=4$m and $L_N=3$m,
while the red line for the one for in the
case with two detectors at $L_F=8$m and $L_N=4$m.
The sensitivity in the case of a conventional
thermal neutron reactor (the green line) is also given
for comparison.}
\label{fig4}
\end{figure}

\subsection{Experimental fast neutron reactors}
In the previous subsection, we have seen that
the sensitivity to $\sin^2{2\theta_{14}}$ is
lost because of the smearing effect of finite
core size.
Fast neutron reactors are known to have
a high power density, and an oscillation
experiments using them might be more
advantageous than those with thermal
neutron reactors.
While experimental fast neutron reactors
have a small power, it may allow experimentalists
to put the detectors at a nearer location because of their
experimental nature.

Here we will discuss the concrete
example of Joyo~\cite{joyo} with MK-III
upgrade~\cite{joyomk3}, which is
an experimental fast breeder reactor.\footnote{
An experiment~\cite{Furuta:2011iu} was performed
to detect neutrinos from a fast neutron reactor
at Joyo, although they did not get
sufficient statistical significance.}
It has a relatively small size
(a diameter of 0.8m, a height of 0.5m),
and a relatively large thermal power 0.14GW.
Its power density is approximately
500 kW/$\ell$, whereas a
typical power density of thermal neutron
reactors is approximately
50 kW/$\ell$.

Again we have computed $\chi^2$ in Eq.~(\ref{chipull})
numerically in the case of Joyo with MK-III upgrade.
Fig.~\ref{fig3} shows the sensitivity
to $\sin^2{2\theta_{14}}$ as a function of
the baseline lengths of the near
and far detectors for $\Delta m^2_{41}=1$eV$^2$,
and 5eV$^2$, respectively.
We have assumed the same conditions for
the detectors (the size, the efficiency,
the exposure time etc.)
as those assumed in Fig.~\ref{fig1}.
From Fig.~\ref{fig3} we see that
the set $(L_N, L_F)\simeq$ (4m, 8m)
gives the optimum for $\Delta m^2_{41}=1$eV$^2$
(Fig.~\ref{fig3} upper panel).

Fig.~\ref{fig4} shows the sensitivity
to $\sin^2{2\theta_{14}}$ as a function of
$\Delta m^2_{41}$ in the case of the sets
of the baseline lengths $(L_N, L_F)=$ (4m, 8m)
and $(L_N, L_F)=$ (3m, 4m).
The set of the baseline lengths $(L_N, L_F)=$ (3m, 4m)
has a peak for $\sin^2{2\theta_{14}}\simeq$ 4 eV$^2$,
while the other set $(L_N, L_F)=$ (4m, 8m)
has sensitivity to $\sin^2{2\theta_{14}}$
for a wider range of $\Delta m^2_{41}$.
The sensitivity in the case of a conventional
thermal neutron reactor is also given in
Fig.~\ref{fig2} for comparison, and
the advantage of the case with Joyo is clear.
Also in this case the statistical errors are
dominant in $\chi^2$, and the result of Fig.~\ref{fig4} 
is almost independent of the number $n$ of bins
for $16\le n\le 64$.

\section{Discussion and Conclusion}

In the framework of the (3+1)-scheme,
we studied the sensitivity to $\sin^22\theta_{14}$
of very short baseline reactor oscillation experiments
by a spectrum analysis.
The assumptions are that we have two detectors
whose size and efficiency are exactly the same as
those used at the Bugey experiment~\cite{Declais:1994su}.

In the case of a conventional thermal neutron reactor,
which has a core of a diameter 4m, a height 4m
and a thermal power 2.8GW,
by putting the detectors at $L_N=$ 17m and
$L_F=$ 23m, we obtain the sensitivity
as good as several $\times 10^{-2}$ for
$\Delta m^2_{41} \lesssim 1$eV$^2$, but
we lose the sensitivity above 1eV$^2$
due to the smearing of the finite core size.

In the case of an experimental fast neutron
reactor with a core of a diameter of 0.8m,
a height of 0.5m and a thermal power 0.14GW,
on the other hand, we obtain the sensitivity
as good as a several $\times 10^{-2}$ for
1eV$^2$ $\lesssim\Delta m^2_{41} \lesssim 10$eV$^2$
if we put the detectors at $L_N=$ 4m and
$L_F=$ 8m.

In both types of reactors with the Bugey-like detector
setup, the statistical errors are dominant, and the
sensitivity is almost independent of the number $n$ of
bins of the spectral analysis.  The reason that the case
of the experimental fast neutron reactor is competitive
despite its small power is because the total numbers of
events at $L\sim$ several meters are comparable to those
of the case with the thermal neutron reactor at $L\sim$ a
few $\times$ 10 meters.

Since the best fit value obtained in
Ref.~\cite{Mention:2011rk} is 
$\Delta m^2_{41} \sim 2$eV$^2$,
an experiment using an experimental fast neutron
reactor offers a promising possibility.

It should be emphasized that
the flux uncertainty by 3\% which was pointed
out in Ref.~\cite{Mueller:2011nm} does not
cause a problem in our discussion.  The reason
is because the main contribution to $\chi^2$
comes from the spectrum analysis, which looks
at the difference of the maximum and the
minimum of the energy spectrum and the effect
of the overall normalization of the flux
(the first term of the RHS in Eq.~(\ref{chi6app}))
gives little contribution to $\chi^2$ in
the region of $\Delta m^2_{41}$ in which
the sensitivity to $\sin^22\theta_{14}$
is better than 0.1.

Although we may not be able to put the detectors at such a
close location, we could increase the detector volume
and/or the measurement period, so that the data size is
the same as those described in this paper.  While the
experimental feasibility of the idea of putting detectors
at a location very near to an experimental fast reactor is
yet to be seen, it is worth investigating whether this
method is technically possible.

\appendix

\section{Derivation of the covariance matrix\label{appendix1}}
By choosing the bins in such a way that $t^A_i=t^A$,
Eq.~(\ref{chipull}) becomes
\begin{eqnarray}
\hspace*{-20mm}
\displaystyle
\chi^2&=&\min_{\alpha's}\Bigg\{
\displaystyle\sum_{A=N,F}\sum_{i=1}^n
\frac{ \left(m^A_i/t^A_i-1-\alpha-\alpha^A-\alpha_i-\alpha_{\text{cal}}^A v^A_i
\right)^2 }{1/t^A + (\sigma^A_{\text{db}})^2 }\nonumber\\
&+&\displaystyle\sum_{A=N,F}\left[\left(
\frac{\alpha^A} {\sigma_{\text{dB}}}\right)^2
+\left(\frac{\alpha_{\text{cal}}^A} {\sigma_{\text{cal}}}\right)^2\right]
+\displaystyle\sum_{i=1}^n\left(
\frac{\alpha_i} {\sigma_{\text{Db}}}\right)^2
+\left(\frac{\alpha} {\sigma_{\text{DB}}}\right)^2\Bigg\}\nonumber\\
&\equiv& \min_{\alpha's} \chi^2_\alpha. \nonumber
\end{eqnarray}
Let us redefine the variables
\begin{eqnarray}
\hspace*{-20mm}
y^A_i&\equiv&\frac{m^A_i-t^A_i} {t^A_i}
= -\sin^22\theta_{14} D^A_i
\nonumber
\end{eqnarray}
and let us introduce a vector notation
\begin{eqnarray}
\hspace*{-20mm}
\vec{y}^{N}\equiv\left(\begin{array}{c}
y^N_1\\
\vdots\\
y^N_n
\end{array}\right),\qquad
\vec{y}^{F}\equiv\left(\begin{array}{c}
y^F_1\\
\vdots\\
y^F_n
\end{array}\right).
\nonumber
\end{eqnarray}
Following the discussions in the Appendix A in~\cite{Sugiyama:2004bv},
the matrix element of the covariance matrix
can be obtained as the expectation value of $y^A_i\,y^B_j$:
\begin{eqnarray}
\hspace*{-20mm}
\left(\rho\right)^{AB}_{ij}&=&
\left\langle y^A_i y^B_j \right\rangle\nonumber\\
&\equiv&{\cal N}
\int d\alpha\prod_{A=N,F}\int d\vec{y}^A\int d\alpha^A\int d\alpha_{\text{cal}}^A
\prod_{i=1}^n\int d\alpha_i~y^A_i\,y^B_j\,
\exp\left(-\frac{\chi^2_\alpha}{2}\right),
\nonumber
\end{eqnarray}
where the normalization ${\cal N}$ is defined in such a way
that $\langle 1 \rangle=1$.  From a straightforward calculation,
we have
\begin{eqnarray}
\hspace*{-20mm}
\left\langle y^A_i y^B_j \right\rangle
&=&\delta^{AB}\delta_{ij}1/t^A
+\delta^{AB}\delta_{ij}\,\sigma_{\text{db}}^2
+\sigma_{\text{DB}}^2
+\delta^{AB}\,\sigma_{\text{dB}}^2
+\delta_{ij}\,\sigma_{\text{Db}}^2
+\delta^{AB}v^A_i v^A_j \,\sigma^2_{\text{cal}}.
\nonumber
\end{eqnarray}
Thus we have
\begin{eqnarray}
\hspace*{-20mm}
\displaystyle
\chi^2=
\left(\begin{array}{ll}
\vec{y}^{N\ T},&\vec{y}^{F\ T}
\end{array}
\right)
\rho^{-1}
\left(\begin{array}{l}
\vec{y}^{N}\\
\vec{y}^{F}
\end{array}\right)
=\sin^42\theta_{14} \left(\begin{array}{ll}
\vec{D}^{N\ T},&\vec{D}^{F\ T}
\end{array}
\right)
\rho^{-1}
\left(\begin{array}{l}
\vec{D}^{N}\\
\vec{D}^{F}
\end{array}\right),
\nonumber
\end{eqnarray}
where the covariance matrix $\rho$ is defined by
\begin{eqnarray}
\hspace*{-20mm}
\rho=\left(\begin{array}{ll}
M+1/t^NI_n&N\\
N&M+1/t^FI_n
\end{array}\right),
\label{rho}
\end{eqnarray}
with
\begin{eqnarray}
\hspace*{-20mm}
M&\equiv& \left(\sigma^2_{\text{Db}}+\sigma^2_{\text{db}}\right)I_n
+\left(\sigma^2_{\text{DB}}+\sigma^2_{\text{dB}}\right)H_n
+\sigma^2_{\text{cal}}G_n,\nonumber\\
N&\equiv& \sigma^2_{\text{Db}}I_n
+\sigma^2_{\text{DB}}H_n.
\label{n}
\end{eqnarray}
Here $I_n$ is an $n\times n$ unit matrix,
$H_n$ and $G_n$ are $n\times n$ matrices defined by
\begin{eqnarray}
\hspace*{-20mm}
H_n&\equiv&\left(\begin{array}{ccc}
1&\cdots&1\\
\vdots&&\vdots\\
1&\cdots&1
\end{array}\right),\label{hn}\\
G_n&\equiv&\left(\begin{array}{cccc}
v_1^2&v_1v_2&\cdots&v_1v_n\\
v_1v_2&v_2^2&\cdots&v_2v_n\\
\vdots&\vdots&&\vdots\\
v_1v_n&v_2v_n&\cdots&v_n^2
\end{array}\right),
\label{gn}
\end{eqnarray}
where $v_j~(j=1,\cdots,n)$ is defined by Eqs.~(\ref{v}),
and (\ref{v3}).
 Note that the covariance matrix does not include
oscillation parameters but only errors.
 Diagonalization of the covariance matrix is useful
to see which errors dominate $\chi^2$.

Unlike in Ref.~\cite{Sugiyama:2005ir},
the presence of the statistical errors $1/t^A$ and
the matrix $G_n$ makes it difficult
to evaluate the eigenvalues analytically, so
only in this appendix, we will take the limit
$\sigma^2_{\text{cal}}\to 0$ which may be justified
in the present setup with the total numbers
of events $\sim$ 10,000 $\times(15\mbox{m}/L_A)^2$.

To diagonalize the matrix (\ref{rho}) with
$\sigma^2_{\text{cal}}=0$,
we first note that the matrix $H_n$ can be
diagonalized as
\begin{eqnarray}
H_n = U_n\,D_n\,U_n^{-1},
\nonumber
\end{eqnarray}
where
\begin{eqnarray}
D_n = \mbox{\rm diag}(n,0,\cdots,0)
\nonumber
\end{eqnarray}
is a diagonal matrix and
\begin{eqnarray}
U_n&=&(\vec{u}_1,\cdots,\vec{u}_n),
\nonumber\\
\vec{u}_1&\equiv&\frac{1}{\sqrt{n}}\,(1,\cdots,1)^T
\nonumber\\
\vec{u}_2&\equiv&\frac{1}{\sqrt{2}}\,(1,-1,0,\cdots,0)^T
\nonumber\\
\vec{u}_3&\equiv&\frac{1}{\sqrt{6}}\,(1,1,-2,0,\cdots,0)^T
\nonumber\\
&{\ }&\cdots
\nonumber\\
\vec{u}_n&\equiv&\frac{1}{\sqrt{n(n-1)}}\,(1,\cdots,1,-(n-1))^T
\nonumber
\end{eqnarray}
is a $n\times n$ unitary matrix.
Thus $M$ and $N$ can be diagonalized as
\begin{eqnarray}
M&=&U_n\,\left\{
\left(\sigma^2_{\text{Db}}+\sigma^2_{\text{db}}\right)I_n
+\left(\sigma^2_{\text{DB}}+\sigma^2_{\text{dB}}\right)D_n
\right\}\,U_n^{-1}
\nonumber\\
&=&U_n\,
\mbox{\rm diag}(p_1,p_2,\cdots,p_2)
\,U_n^{-1},
\nonumber\\
N&=&U_n\,\left(
\sigma^2_{\text{Db}}I_n
+\sigma^2_{\text{DB}}D_n
\right)\,U_n^{-1}
\nonumber\\
&=&U_n\,
\mbox{\rm diag}(q_1,q_2,\cdots,q_2)
\,U_n^{-1},
\nonumber
\end{eqnarray}
where
\begin{eqnarray}
p_1&\equiv&\sigma^2_{\text{Db}}+\sigma^2_{\text{db}}
+n(\sigma^2_{\text{DB}}+\sigma^2_{\text{dB}})
\nonumber\\
p_2&\equiv&\sigma^2_{\text{Db}}+\sigma^2_{\text{db}}
\nonumber\\
q_1&\equiv&\sigma^2_{\text{Db}}
+n\sigma^2_{\text{DB}}
\nonumber\\
q_2&\equiv&\sigma^2_{\text{Db}}.
\label{pq}
\end{eqnarray}

Now (\ref{rho}) can be cast into a block diagonal by
\begin{eqnarray}
\hspace*{-20mm}
&{\ }&
\left(\begin{array}{cc}
U_n^{-1}&0\\
0&U_n^{-1}
\end{array}\right)
\left(\begin{array}{cc}
M+1/t^NI_n&N\\
N&M+1/t^FI_n
\end{array}\right)
\left(\begin{array}{cc}
U_n&0\\
0&U_n
\end{array}\right)
\nonumber\\
&=&
\left(\begin{array}{cc}
\left(\sigma^2_{\text{Db}}+\sigma^2_{\text{db}}+1/t^N\right)I_n
+\left(\sigma^2_{\text{DB}}+\sigma^2_{\text{dB}}\right)D_n
&\sigma^2_{\text{Db}}I_n
+\sigma^2_{\text{DB}}D_n\\
\sigma^2_{\text{Db}}I_n
+\sigma^2_{\text{DB}}D_n&
\left(\sigma^2_{\text{Db}}+\sigma^2_{\text{db}}+1/t^F\right)I_n
+\left(\sigma^2_{\text{DB}}+\sigma^2_{\text{dB}}\right)D_n
\end{array}\right)
\nonumber\\
&=&
{\cal U}
\left(\begin{array}{ccccc}
B_1
&0&\cdots&&0\\
0&
B_2
&\ddots&&\vdots\\
\vdots&\ddots&\ddots&&0\\
0&\cdots&0&&
B_2
\end{array}\right)
{\cal U}^T,
\nonumber
\end{eqnarray}
where ${\cal U}$ is a $2n\times 2n$ unitary
matrix whose matrix elements are given by
\begin{eqnarray}
{\cal U}_{pq}
\equiv
\left(\begin{array}{cccc|cccc}
1&0&\cdots&0&0&0&\cdots&0\\
0&0&\cdots&0&1&0&\cdots&0\\
0&1&0&\cdots&0&0&\cdots&0\\
0&0&\ddots&0&0&1&\cdots&0\\
0&&\cdots&1&&&\ddots&0\\
0&0&\cdots&0&0&\cdots&0&1
\end{array}\right)_{pq}
=\sum_{j=1}^n\left(
\delta_{p, 2j-1}\delta_{q, j}
+\delta_{p, 2j}\delta_{q, n+j}\right).
\nonumber
\end{eqnarray}
$B_1$, $B_2$ are $2\times2$ matrices
defined as follows and are diagonalized as
\begin{eqnarray}
B_1&\equiv&\left(\begin{array}{cc}
p_1+1/t^N&q_1\\
q_1&p_1+1/t^F
\end{array}\right)
=
e^{i\varphi_1\sigma_2}\,
\mbox{\rm diag}(\lambda_1^{(-)},\lambda_1^{(+)})
\,e^{-i\varphi_1\sigma_2}
\nonumber\\
B_2&\equiv&\left(\begin{array}{cc}
p_2+1/t^N&q_2\\
q_2&p_2+1/t^F
\end{array}\right)
=
e^{i\varphi_2\sigma_2}\,
\mbox{\rm diag}(\lambda_2^{(-)},\lambda_2^{(+)})
\,e^{-i\varphi_2\sigma_2},
\nonumber
\end{eqnarray}
where
\begin{eqnarray}
\sigma_2\equiv\left(\begin{array}{cc}
0&-i\\
i&0
\end{array}\right)
\nonumber
\end{eqnarray}
is the Pauli matrix,
$\varphi_1$ and $\varphi_2$ are introduced to
diagonalize each
matrix and are given by
\begin{eqnarray}
\tan2\varphi_1&=&\frac{2q_1}{1/t^F-1/t^N}
=\frac{2(\sigma^2_{\text{Db}}
+n\sigma^2_{\text{DB}})}{1/t^F-1/t^N}
\nonumber\\
\tan2\varphi_2&=&\frac{2q_2}{1/t^F-1/t^N}
=\frac{2\sigma^2_{\text{Db}}}{1/t^F-1/t^N},
\label{phi}
\end{eqnarray}
and $\lambda^{(\pm)}_j~(j=1,2)$ are the
eigenvalues given by
\begin{eqnarray}
\lambda^{(\pm)}_1&\equiv&
n\left[
\frac{\sigma^2_{\text{Db}}+\sigma^2_{\text{db}}}{n}
+\sigma^2_{\text{DB}}+\sigma^2_{\text{dB}}
+\frac{1}{2}\left(\frac{1}{N^F}+\frac{1}{N^N}\right)
\right.
\nonumber\\
&{\ }&\pm\left.\left\{
\frac{1}{4}\left(\frac{1}{N^F}-\frac{1}{N^N}\right)^2
+\left(\frac{\sigma^2_{\text{Db}}}{n}
+\sigma^2_{\text{DB}}\right)^2
\right\}\right]
\nonumber\\
\lambda^{(\pm)}_2&\equiv&
n\left[
\frac{\sigma^2_{\text{Db}}+\sigma^2_{\text{db}}}{n}
+\frac{1}{2}\left(\frac{1}{N^F}+\frac{1}{N^N}\right)
\right.
\nonumber\\
&{\ }&\pm\left.\left\{
\frac{1}{4}\left(\frac{1}{N^F}-\frac{1}{N^N}\right)^2
+\left(\frac{\sigma^2_{\text{Db}}}{n}
\right)^2
\right\}\right].
\nonumber
\end{eqnarray}
In the last expression, we have factored out $n$ from the
eigenvalues for later convenience, and
we have introduced the total numbers of events
$N^A\equiv n\times t^A~(A=N,F)$.

To evaluate $\chi^2$, we need to calculate
the following factor:
\begin{eqnarray}
&{\ }&
\left(\begin{array}{cccc}
e^{-i\varphi_1\sigma_2}
&0&\cdots&0\\
0&e^{-i\varphi_2\sigma_2}&\cdots&0\\
&&\ddots&\\
0&\cdots&&e^{-i\varphi_2\sigma_2}
\end{array}\right)
{\cal U}
\left(\begin{array}{cc}
U_n^{-1}&0\\
0&U_n^{-1}
\end{array}\right)
\left(\begin{array}{c}
\vec{D}^{N}\\
\vec{D}^{F}
\end{array}\right)
\nonumber\\
&=&
\left(\begin{array}{cccc}
e^{-i\varphi_1\sigma_2}
&0&\cdots&0\\
0&e^{-i\varphi_2\sigma_2}&\cdots&0\\
&&\ddots&\\
0&\cdots&&e^{-i\varphi_2\sigma_2}
\end{array}\right)
{\cal U}
\left(\begin{array}{c}
\vec{u}_1\cdot\vec{D}^{N}\\
\vdots\\
\vec{u}_n\cdot\vec{D}^{N}\\
\vec{u}_1\cdot\vec{D}^{F}\\
\vdots\\
\vec{u}_n\cdot\vec{D}^{F}
\end{array}\right)
\nonumber\\
&=&
\left(\begin{array}{c}
e^{-i\varphi_1\sigma_2}
\left(\begin{array}{c}
\vec{u}_1\cdot\vec{D}^{N}\\
\vec{u}_1\cdot\vec{D}^{F}
\end{array}\right)\\
e^{-i\varphi_2\sigma_2}
\left(\begin{array}{c}
\vec{u}_2\cdot\vec{D}^{N}\\
\vec{u}_2\cdot\vec{D}^{F}
\end{array}\right)\\
\vdots\\
e^{-i\varphi_2\sigma_2}
\left(\begin{array}{c}
\vec{u}_n\cdot\vec{D}^{N}\\
\vec{u}_n\cdot\vec{D}^{F}
\end{array}\right)
\end{array}\right)
=
\left(\begin{array}{c}
\left(\begin{array}{c}
c_1\vec{u}_1\cdot\vec{D}^{N}
-s_1\vec{u}_1\cdot\vec{D}^{F}\\
c_1\vec{u}_1\cdot\vec{D}^{F}
+s_1\vec{u}_1\cdot\vec{D}^{N}
\end{array}\right)\\
\left(\begin{array}{c}
c_2\vec{u}_2\cdot\vec{D}^{N}
-s_2\vec{u}_2\cdot\vec{D}^{F}\\
c_2\vec{u}_2\cdot\vec{D}^{F}
+s_2\vec{u}_2\cdot\vec{D}^{N}
\end{array}\right)\\
\vdots\\
\left(\begin{array}{c}
c_2\vec{u}_n\cdot\vec{D}^{N}
-s_2\vec{u}_n\cdot\vec{D}^{F}\\
c_2\vec{u}_n\cdot\vec{D}^{F}
+s_2\vec{u}_n\cdot\vec{D}^{N}
\end{array}\right)
\end{array}\right),
\nonumber
\end{eqnarray}
where $c_j\equiv\cos\varphi_j,
s_j\equiv\sin\varphi_j~(j=1,2)$.

Putting everything together, we obtain
\begin{eqnarray}
\hspace*{-35mm}
&{\ }&\frac{\chi^2}{\sin^42\theta_{14}}
\nonumber\\
&=&
\frac{\left(c_1\vec{u}_1\cdot\vec{D}^{F}
+s_1\vec{u}_1\cdot\vec{D}^{N}\right)^2/n}
{\sigma^2_{\text{DB}}+\sigma^2_{\text{dB}}
+(\sigma^2_{\text{Db}}+\sigma^2_{\text{db}})/n
+(1/N^F+1/N^N)/2
+\left\{(1/N^F-1/N^N)^2/4
+(\sigma^2_{\text{DB}}+\sigma^2_{\text{Db}}/n)^2
\right\}^{1/2}}
\nonumber\\
&{\ }&
+\frac{
\left(c_1\vec{u}_1\cdot\vec{D}^{N}
-s_1\vec{u}_1\cdot\vec{D}^{F}\right)^2/n}
{\sigma^2_{\text{DB}}+\sigma^2_{\text{dB}}
+(\sigma^2_{\text{Db}}+\sigma^2_{\text{db}})/n
+(1/N^F+1/N^N)/2
-\left\{(1/N^F-1/N^N)^2/4
+(\sigma^2_{\text{DB}}+\sigma^2_{\text{Db}}/n)^2
\right\}^{1/2}
}
\nonumber\\
&{\ }&
+\sum_{j=2}^n\frac{
\left(c_2\vec{u}_j\cdot\vec{D}^{F}
+s_2\vec{u}_j\cdot\vec{D}^{N}\right)^2/n}
{(\sigma^2_{\text{Db}}+\sigma^2_{\text{db}})/n
+(1/N^F+1/N^N)/2
+\left\{(1/N^F-1/N^N)^2/4
+(\sigma^2_{\text{Db}}/n)^2
\right\}^{1/2}
}
\nonumber\\
&{\ }&
+\sum_{j=2}^n\frac{
\left(c_2\vec{u}_j\cdot\vec{D}^{N}
-s_2\vec{u}_j\cdot\vec{D}^{F}\right)^2/n}
{(\sigma^2_{\text{Db}}+\sigma^2_{\text{db}})/n
+(1/N^F+1/N^N)/2
-\left\{(1/N^F-1/N^N)^2/4
+(\sigma^2_{\text{Db}}/n)^2
\right\}^{1/2}
},
\label{chi6}
\end{eqnarray}
where both the numerators and denominators are divided by
the number $n$ of bins, because the numerators are almost
independent of $n$ after being divided by $n$ in the limit
of infinite statistics~\cite{Sugiyama:2005ir}.

In the present case,
assuming that the total numbers of events
$N^F$ and $N^N$ are of order $10^5$ and that
the number $n$ of bins is 32,
our reference values (\ref{error}) satisfy
$\sigma^2_{\text{DB}}\simeq 9\times 10^{-4}\gg
1/N^A\sim 10^{-5}~(A=N,F)$,
$\sigma^2_{\text{dB}}\simeq 3\times10^{-5}\gg
\sigma^2_{\text{db}}/n\simeq 1\times10^{-6}$,
$\sigma^2_{\text{Db}}/n\simeq 1\times10^{-5}
\sim 1/N^A\sim 10^{-5}~(A=N,F)$.
Hence we obtain Eq.~(\ref{chi6app}).

\begin{acknowledgments}
The author would like to thank F.~Suekane
for informing him of Ref.~\cite{joyomk3}
and for useful correspondence.
This work was partly supported by Grants-in-Aid for Scientific Research
of the Ministry of Education, Culture, Sports, Science, and Technology,
under Grant No.\ 21540274.
\end{acknowledgments}


\end{document}